\documentclass[runningheads]{llncs}

\usepackage[hidelinks]{hyperref}

\usepackage[utf8]{inputenc}
\usepackage{xcolor}
\usepackage{xspace}
\usepackage{amsmath}
\usepackage{mathpartir}
\usepackage{listings}

\usepackage{enumitem}
\usepackage{tikz}
\usetikzlibrary{arrows, decorations.pathreplacing}

\definecolor{pblue}{rgb}{0.13,0.13,1}
\definecolor{pgreen}{rgb}{0,0.5,0}
\definecolor{pgrey}{rgb}{0.46,0.45,0.48}
\definecolor{pwhite}{rgb}{1.0,1.0,1.0}
\lstdefinelanguage{myML}{
  morekeywords={match,with,proc,assert,for,foreach,range,do,while,until,break,if,then,else,return,and,or,raise,false,true},
  columns=fullflexible,
  sensitive=true,
  commentstyle = \itshape, 
  morecomment={[l]//},
  mathescape=true,
  basicstyle=\small,
  identifierstyle={\ttfamily},
  stringstyle=\rmfamily,
  keywordstyle=\color{black},
  keywordstyle=\bfseries,
  literate={<-}{$\leftarrow\ $}{2} {->}{$\rightarrow\ $}{2} {:=}{$\leftarrow\ $}{2}
}

\newcommand{\Var}{\mathit{Var}}
\newcommand{\Fld}{\mathit{Fld}}

\newcommand{\Cls}{\mathit{Cls}}
\newcommand{\Mtd}{\mathit{Mtd}}
\newcommand{\Expr}{\mathit{Expr}}
\newcommand{\Pos}{\mathit{Pos}}
\newcommand{\Reg}{\mathit{Reg}}
\newcommand{\Typ}{\mathit{Typ}}

\newcommand{\Object}{\texttt{Object}\xspace}
\newcommand{\NullType}{\texttt{NullType}\xspace}

\newcommand{\itnull}{\mathit{null}}
\newcommand{\this}{\texttt{this}}
\newcommand{\letin}[3]{\texttt{let} \; #1=#2 \; \texttt{in} \; #3}
\newcommand{\ttifthenelse}[4]{\texttt{if} \; #1=#2 \; \texttt{then} \; #3 \; \texttt{else} \; #4}

\newcommand{\ttnull}{\texttt{null}}
\newcommand{\new}[2]{\texttt{new}^{#1}\; #2}
\newcommand{\cast}[2]{(#1)\, #2}
\newcommand{\tto}{\texttt{emit}}

\newcommand{\fields}{\mathit{fields}}
\newcommand{\methods}{\mathit{methods}}
\newcommand{\mtable}{\mathit{mtable}}

\newcommand{\cP}{\mathcal{P}}
\newcommand{\cPf}{\cP^{\text{fin}}}

\newcommand{\dom}{\mathrm{dom}}

\newcommand{\cS}{\mathcal{S}}

\newcommand{\cR}{\mathcal{R}}

\newcommand{\Node}{\texttt{Node}\xspace}
\newcommand{\ttlast}{\texttt{last}\xspace}

\newcommand{\cyclic}{\texttt{cyclic}\xspace}
\newcommand{\linear}{\texttt{linear}\xspace}
\newcommand{\ttnext}{\texttt{next}\xspace}

\newcommand{\Null}{\texttt{Null}}
\newcommand{\Createdat}[1]{\texttt{CreatedAt}\mathopen{}\left( #1 \right)\mathclose{}}

\newcommand{\of}{{:}\,}
\newcommand{\Mid}{\,\mid\,}

\newcommand{\env}{\mathit{env}}
\newcommand{\args}{\mathit{args}}
\newcommand{\sub}[1]{[\![#1]\!]}

\newcommand{\ATrans}{\textit{ATrans}}
\newcommand{\Tm}{\textit{Tm}}
\newcommand{\fgraph}[3]{\langle #1,#2,#3 \rangle}
\newcommand{\FG}{\mathcal{F}}
\newcommand{\efgraph}{\mathcal{E}}
\newcommand{\FK}{\mathcal{K}}

\newcommand{\tgeq}{:\geq}
\newcommand{\tmapsto}{:\mapsto}
\newcommand{\vpart}[1]{#1|_\mathrm{v}}
\newcommand{\fpart}[1]{#1|_\mathrm{f}}

\begin{document}

\title{Inferring Region Types via an Abstract Notion of Environment Transformation\thanks{Supported by the German Research Foundation (DFG) under the research grant 250888164 (GuideForce).}}
\titlerunning{Abstract Transformations}

\author{Ulrich Sch\"opp\orcidID{0000-0002-5445-9461} \and
 Chuangjie Xu\orcidID{0000-0001-6838-4221}}

\authorrunning{U.\ Sch\"opp \and C.\ Xu}

\institute{fortiss GmbH, Guerickestraße 25, 80805 Munich, Germany}

\maketitle

\begin{abstract}
Region-based type systems are a powerful tool for various kinds of program analysis. We introduce a new inference algorithm for region types based on an abstract notion of environment transformation. It analyzes the code of a method only once, even when there are multiple invocations of the method of different region types in the program. Elements of such an abstract transformation are essentially constraints for equality and subtyping that capture flow information of the program. In particular, we work with access graphs in the definition of abstract transformations to guarantee the termination of the inference algorithm, because they provide a finite representation of field access paths.

\end{abstract}

\section{Introduction}

Programs typically make extensive use of libraries. Analyzing a program thus often involves analysis of big libraries which can be heavy and expensive. The situation gets worse for those analyses where multiple invocations of the same library method requires to re-analyze the library. Therefore, it is significant for analyses to be \emph{compositional}, that is, the analysis result of a program can be computed from the results of its components. Once a library has been analyzed, the result can be directly used to analyze programs that use the library. This work aims at making \emph{region type inference} compositional.

Region-based type systems have been illustrated to be a powerful tool for e.g.\ memory management~\cite{10.1145/781131.781168,10.1145/996841.996871}, pointer analysis and taint analysis~\cite{BGH:pointer,GHL:enforcing,10.1007/978-3-642-54804-8_10}. The usage of regions in effect-and-type systems can improve the precision of analysis of trace properties~\cite{EHZ:enforcing,ESX:infinitary}. The idea of these type-based analysis approaches are to infer the type of a program which allows one to verify if the program satisfies certain properties. However, the type inference algorithms for the region type systems for Featherweight Java from the previous work~\cite{BGH:pointer,EHZ:enforcing,ESX:infinitary,GHL:enforcing} are \emph{not} compositional. The type of a method is inferred from the ones of its arguments. If the method is called with arguments of different types, its code is analyzed multiple times, one for each invocation.

To avoid redundant analysis, we introduce a new inference algorithm based on an abstract notion of environment transformation. The idea is to summarize the flow information of the program using an abstract transformation. Then we derive the type of a method by applying its abstract transformation to the types of its arguments. When analyzing some new code which invokes some methods that have been analyzed, we can use the abstract transformations computed in the previous round of analysis, rather than re-analyzing the code of the methods as in the previous work~\cite{BGH:pointer,EHZ:enforcing,ESX:infinitary,GHL:enforcing}. We now explain the idea in more detail.

\paragraph{Region Types and Typing Environments.}
We work with the region type system of Beringer et al.~\cite{BGH:pointer} for Featherweight Java~\cite{DBLP:journals/toplas/IgarashiPW01}. But our approach can be adapted for other type systems. In our type system, \emph{region types} represent some properties of values. For example, we consider a region $\Createdat{\ell}$ for references to objects that were created in the position with label~$\ell$. One can think of the label $\ell$ as a line number in the source code. This region enables us to track where in the program an object originates. We allow \emph{typing environments} to carry field typing. For example, the environment
\[
E = (x : \Createdat{\ell_1},\ \Createdat{\ell_1}.f : \Createdat{\ell_2})
\]
means that $x$ points to an object which is created at position $\ell_1$ and the field $f$ of any object created at $\ell_1$ is an object created at~$\ell_2$.

\paragraph{Environment Transformations.}
Inferring region types is essentially a flow analysis. The execution of a program may change the types of its variables and fields. Thus we want to assign it an \emph{environment transformation} that captures how the types are updated in the program. For example, the program
\[
\begin{array}{l}
y = x.f;\\
x = \new{\ell_3}{C()};
\end{array}
\]
can be assigned the transformation
\[
[y \tmapsto x.f,\ x \tmapsto \Createdat{\ell_3}].
\]
It updates the environment $E$ to
\[
(x : \Createdat{\ell_3},\ y : \Createdat{\ell_2},\ \Createdat{\ell_1}.f : \Createdat{\ell_2}).
\]
Note that the substitutions are performed simultaneously. If the program returns the variable $x$, then we look it up in the above updated environment and conclude that the program has return type $\Createdat{\ell_3}$, meaning that it returns an object created at position $\ell_3$.

\paragraph{Field Access Graphs.}
Directly using field access paths like $x.f$ in environments as above is problematic, because the lengths of access paths may be unbounded. The computation of environment transformations involving such access paths may not terminate. For example, consider a class of linked lists with a field $\ttnext : \Node$ pointing to the next node. The following method returns the last node of a list.
\[
\begin{array}{l}
\Node \ \ttlast() \ \{ \\
\quad \mathtt{if} \ (\ttnext \ \mathtt{==} \ \ttnull) \ \{ \mathtt{return} \ \this; \} \\
\quad \mathtt{else} \ \{ \mathtt{return} \ \ttnext.\ttlast(); \} \\
\}
\end{array}
\]
Its return type can be the same as the type of variable $\this$, or the types of the paths $\this.\ttnext$, $\this.\ttnext.\ttnext$ and so on, resulting in an infinite set of access paths. To solve this, we work with \emph{access graphs} which provide a finite representation of access paths~\cite{10.1145/1290520.1290521,spth_et_al:LIPIcs:2016:6116}. For example, the $\Node$ class has three access graphs to represent all its access paths. The return type of $\ttlast$ is then computed via the set containing these three graphs. 

\paragraph{Field Update and Constraints.}
We work with \emph{weak update} for field typing as in~\cite{BGH:pointer}: If a field $f$ of some object is assigned a value of type $B$, and in another occasion it is assigned a value of type $C$, then the field should have a type containing both $B$ and $C$. Therefore, for an assignment statement like $y.f = x$, we assign it a \emph{constraint} $y.f \tgeq x$, meaning that the type of the field $f$ of any object of the type of $y$ should be greater than or equal to the type of $x$.

\paragraph{Abstract Transformations.}
With the above ingredients, we introduce a notion of abstract transformation. An \emph{abstract transformation} consists of assignments $x \tmapsto u$ and constraints $\kappa \tgeq v$. The value such as $u,v$ is a formal disjunction of some atoms. An \emph{atom} is a variable, a type or a field graph following a variable or a type. The key $\kappa$ is a non-empty graph representing access paths. To capture how types are updated in a program, we define the following operations on abstract transformations. We \emph{instantiate} an abstract transformation to an endofunction on typing environments. It computes the types of variables and fields of a program with a given initial typing. We define the \emph{composition} of abstract transformations to model type updates in a statement followed by another. We also define the \emph{join} of abstract transformations to tackle conditional branches.

\paragraph{Type Inference.}
Suppose we have a table~$T$ assigning an abstract transformation to each method of a program. Then we can compute an abstract transformation for any expression~$e$ of the program by induction on~$e$. For example, when~$e$ is an invocation of a method, we lookup the table $T$ to get the abstract transformation; and when~$e$ is a conditional expression, we join the abstract transformations of its branches. For any well-typed program, we have a fixed-point algorithm to compute such a table~$T$ for it. To infer the type of a method, we find its abstract transformation from $T$, feed it with the argument types, and then get the type of the return variable from the resulting typing environment.

\subsubsection{Related Work}
Constraint-based analysis is a common technique for type inference with a rich history~\cite{10.1007/3-540-58485-4_34,10.1007/3-540-19027-9_7,10.1145/165180.165188,OSW:constraint,10.1145/118014.117965,10.1007/978-3-642-25318-8_10,10.1145/99370.99394,10.5555/646158.679874}. It may be divided into two main phases. The first phase is to generate constraints by traversing the program. To improve the efficiency of type inference, some simplification may be performed on the generated constraints. Our computation of the table $T$ of abstract transformations corresponds to constraint generation, where constraints are simplified by the composition operation of abstract transformations. The second phase is to solve the generated constraints. There are many different constraint solvers. In our approach, we instantiate the abstract transformations in~$T$ to infer the type of the program, which corresponds to constraint solving. Therefore, our approach is essentially a constraint-based type inference algorithm. But it departs from the existing work in the following aspects. We make use of constraints to infer region information of the program rather than implementation types (i.e., sets of classes)~\cite{10.1007/3-540-58485-4_34,10.1145/118014.117965,10.5555/646158.679874}. Moreover, we work with access graphs for the constraint language to guarantee the termination of our inference algorithm, rather than requiring an additional termination test~\cite{10.1007/978-3-642-25318-8_10,10.5555/646158.679874}.

Our approach is also closely related to the framework for Interprocedural Distributive Environments (IDE) of Sagiv et al.~\cite{10.1016/0304-3975(96)00072-2}. The main idea of the IDE framework is to reduce a program-analysis problem to a pure graph-reachability problem. A user defines a set of environment transformers, that is, endofunctions on environments describing the effect of a statement, and then uses an IDE solver such as Heros~\cite{10.1145/2259051.2259052,heros} to compute analysis results for the entire program. In particular, IDE requires environment transformers to be distributive: transforming the join of any environments gives the same result of joining the transformed environments. We attempted to use IDE to infer region types, but the environment transformer for statement such as $x=y.f$ is not distributive, because it needs to access the input environment multiple times in order to get the type of $y.f$. This failed attempt motivated us to develop a symbolic representation of environment transformers for type inference, resulting in our notion of abstract transformation.


\section{Background}
\label{sec:background}

We briefly recall the definitions of Featherweight Java and access graphs.

\subsection{Featherweight Java}
\label{sec:FJ}
We work with a variant of Featherweight Java (FJ) using the formulation of~\cite{EHZ:enforcing}. It extends FJ~\cite{DBLP:journals/toplas/IgarashiPW01} with field updates, and has primitive if- and let-expressions for convenience. In the presence of field updates, we omit constructors for simplicity. 

The syntax of the language uses four kinds of names.
\[
\begin{array}{rlcrl}
\text{variables: } & x,y \in \Var
& \qquad &
\text{classes: } & C,D \in \Cls
\\
\text{fields: } & f \in \Fld
& &
\text{methods: } & m \in \Mtd
\end{array}
\]
Program expressions are defined as follows:
\[
\begin{aligned}
\Expr \ni e ::=  \ x &\Mid \letin{x}{e_1}{e_2} \Mid \ttifthenelse{x}{y}{e_1}{e_2} \\
& \Mid \ttnull \Mid \new{\ell}{C} \Mid \cast{C}{e} \Mid x^C.m(\bar{y}) \Mid x^C.f \Mid x^C.f := y
\end{aligned}
\]
The expression $\new{\ell}{C}$ creates a new object of class $C$ with all fields initiated to $\ttnull$. It is annotated with a label~$\ell \in \Pos$. We use labels only to distinguish different occurrences of $\mathtt{new}$ in a program, since our type system will track where objects were created. In a few expressions we have added type annotations and write $x^C$ for a variable of class~$C$. They will be needed when looking up in the class table. This is simpler than working with typed variable declarations, since we do not need to find the declarations in order to get the type of a variable. We sometimes omit annotations when they are not needed.

We assume three distinguished formal elements: $\Object, \NullType \in \Cls$ and $\this \in \Var$. The $\NullType$ class plays the role of the type of $\ttnull$ from the Java language specification~\cite[\S4]{JavaSpec}. It may not be used in programs, i.e.~we require $C \not= \NullType$ in create expression $\new{\ell}{C}$ and casting expression $\cast{C}{e}$. When $x$ is not a free variable of $e_2$, we may write $e_1;e_2$ rather than $\letin{x}{e_1}{e_2}$.

An FJ program $(\prec,\fields,\methods,\mtable)$ consists of
\begin{itemize}
\item a subtyping relation ${\prec} \in \cPf(\Cls \times \Cls)$ with $C \prec D$ meaning that $C$ is an immediate subclass of $D$,
\item a field list $\fields: \Cls \to \cPf(\Fld)$ mapping a class to its fields,
\item a method list $\methods: \Cls \to \cPf(\Mtd)$ mapping a class to its methods,
\item a method table $\mtable: \Cls \times \Mtd \rightharpoonup \Var^* \times \Expr$ mapping a method to the pair of its formal parameters and its body.
\end{itemize}
All components are required to be well-formed.  We refer the reader to e.g.~\cite[Section~3]{EHZ:enforcing} for  details. Let $\preceq$ be the reflexive and transitive closure of $\prec$. Then we have $C\preceq \Object$ and $\NullType \preceq C$ for any class $C\in \Cls$.

In the standard FJ type system~\cite{DBLP:journals/toplas/IgarashiPW01}, types are simply classes. In the rest of this paper, we consider only FJ programs that are well-typed with respect to the standard FJ type system.

\subsection{Access Graphs}
\label{sec:access:graph}

For recursive data types such as linked lists, the lengths of access paths may be unbounded. If environment transformations are defined upon access paths, their computation may not terminate. In this paper, we choose to work with the finite representation of access paths given by access graphs~\cite{10.1145/1290520.1290521,spth_et_al:LIPIcs:2016:6116} among the others~\cite{10.1145/773473.178263,10.1145/567752.567776,10.1109/ASE.2015.9}.

An \emph{access graph} $x.G$ consists of a local variable~$x$, called its \emph{base}, and a field graph $G$. A \emph{field graph} is a directed graph whose nodes are fields. The empty field graph is denoted by $\efgraph$. The access graph $x.\efgraph$ represents the plain variable~$x$. Thus we often omit the empty field graph $\efgraph$ and simply write~$x$. If a field graph is not empty, it has a head node $h \in \Fld$ and a tail node $t \in \Fld$ such that for each node $n \in \Fld$ within the field graph there exists a path from~$h$ to~$t$ passing through~$n$. Note that the head and tail can be the same. A non-empty field graph can be uniquely identified by its head~$h$, tail~$t$ and edge set $E \subseteq \Fld \times \Fld$; thus, we write $\fgraph{h}{E}{t}$ to denote it. Each access graph $x.\fgraph{h}{E}{t}$ represents the set of access paths obtained by traversing the field graph from the head to the tail.
We write $\FG$ to denote the set of field graphs and use $G,G'$ to range over field graphs in the paper.

\begin{example}
Consider the following access graphs for a class of nodes for linked lists. The field $v$ is the value stored in the current node and the field $n$ points to the next node.
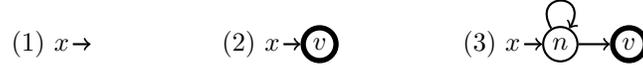
\begin{center}
\begin{tikzpicture}[node distance={26pt}, thick, main/.style = {draw, circle, inner sep=2pt}, var/.style = {inner sep=1pt}]
\node[var] (x1) at(0,0) {\ref{x1} $x$};
\draw[->] (x1) -- (0.67,0);
\node[var] (x2) at(2.8,0) {\ref{x2} $x$};
\node[main, line width=2pt] (v2) [right of=x2] {$v$};
\draw[->] (x2) -- (v2);
\node[var] (x3) at(6,0) {\ref{x3} $x$};
\node[main] (n3) [right of=x3] {$n$};
\node[main, line width=2pt] (v3) [right of=n3] {$v$};
\draw[->] (x3) -- (n3);
\draw[->] (n3) -- (v3);
\draw[->] (n3) to [out=120,in=60,looseness=6] (n3);
\end{tikzpicture}
\end{center}
In the above diagrams, each bold circle represents a tail. These access graphs represent access paths as explained below:
\begin{enumerate}[label={(\arabic*)}]
\item \label{x1} $x.\efgraph$ represents the variable $x$.
\item \label{x2} $x.\fgraph{v}{\emptyset}{v}$ represents the path $x.v$.
\item \label{x3} $x.\fgraph{n}{\{(n,n),(n,v)\}}{v}$ represents the paths $x.n.v$, $x.n.n.v$ and so on. \qed
\end{enumerate}
\end{example}

Given any two field graphs $G$ and $G'$, we \emph{concatenate} them and obtain a field graph $G.G' \in \FG$ as follows:
 \[
 \begin{aligned}
 G.\efgraph & := G \\
 \efgraph.G' & := G' \\
 \fgraph{h}{E}{t}.\fgraph{h'}{E'}{t'} & := \fgraph{h}{E \cup \{ (t,h') \} \cup E'}{t'}.
 \end{aligned}
 \]
Intuitively, the concatenation of a path in $G$ with one in $G'$ lives in $G.G'$. This operation is needed for defining composition of environment transformations.

We work with a generalization of access graphs $b.G$ where $b$ can be either a variable or a type in order to model field typing as explained in Session~\ref{sec:trans:def}.

\section{A Theory of Abstract Transformations}

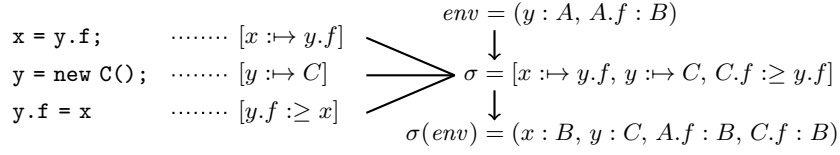
\begin{figure}[h!]
\begin{center}
\begin{tikzpicture}[node distance={5pt}, thick,
    code/.style  = {anchor = west, text width = 57pt, text height = 6pt},
    trans/.style = {code, text width = 45pt},
    env/.style   = {code, text width = 142pt}]
\node[code] (l1) at(0,0)    {\verb|x = y.f;|};
\node[code] (l2) at(0,-0.5) {\verb|y = new C();|};
\node[code] (l3) at(0,-1)   {\verb|y.f = x|};
\node[trans] (t1) at (3,0)    {$[x \tmapsto y.f]$};
\node[trans] (t2) at (3,-0.5) {$[y \tmapsto C]$};
\node[trans] (t3) at (3,-1)   {$[y.f \tgeq x]$};
\draw[dotted] (l1) -- (t1);
\draw[dotted] (l2) -- (t2);
\draw[dotted] (l3) -- (t3);
\node[env] (t) at (6,-0.5)
  {$\sigma = 
    [x \tmapsto y.f,\,
     y \tmapsto C,\,
     C.f \tgeq y.f]$};
\draw (t1) to [out=0,in=180,looseness=0] (t);
\draw (t2) to [out=0,in=180,looseness=0] (t);
\draw (t3) to [out=0,in=180,looseness=0] (t);
\node[env] (e1) at (5.7,0.3)
  {$\env = 
    (y : A,\,
     A.f : B)$};
\node[env, text width = 165pt] (e2) at (5.23,-1.3)
  {$\sigma(\env) = 
    (x : B,\,
     y : C,\,
     A.f : B,\,
     C.f : B)$};
\draw[->] (6.53,0.1) -- (6.53,-0.3);
\draw[->] (6.53,-0.7) -- (6.53,-1.1);
\end{tikzpicture}
\end{center}
\caption{An example illustrating the idea of typing via environment transformations}
\label{fig:code:trans:env}
\end{figure}

Our idea is to type a program via environment transformations. Consider the simple example given in Fig.~\ref{fig:code:trans:env}. Each statement of the program is assigned an environment transformation. They are composed into an environment transformation~$\sigma$ for the whole program. For any given initial typing environment $\env$, we obtain the updated environment $\sigma(\env)$ containing the typing information after executing the program. Lastly, we get the return type of the program from the updated typing environment $\sigma(\env)$. In this section, we explain what environment transformations are and how they update typing environments.

This section is organized as follows. Section~\ref{sec:type:env} presents the assumptions and definitions of types and typing environments. Section~\ref{sec:trans:def} introduces our abstract notion of environment transformation which is based on access graphs. Lastly, Section~\ref{sec:trans:ops} demonstrates some operations on abstract transformations which are essential for modeling the type updates of the program.

\subsection{Types and Environments}
\label{sec:type:env}

We use abstract transformations to encode the changes of types in the program. But our approach is general and works for various type systems including those in the previous work~\cite{BGH:pointer,GHL:enforcing,EHZ:enforcing,ESX:infinitary}. We target at flow type systems in the spirit of Microsoft's TypeScript~\cite{typescript} and Facebook's Flow~\cite{flow}, rather than the standard FJ typing~\cite{DBLP:journals/toplas/IgarashiPW01}. We leave the notion of type generic in this section. For instance, when working with classes, our approach can infer implementation types~\cite{10.1007/3-540-58485-4_34,10.1145/118014.117965,10.5555/646158.679874}. In the next section, we work with region types to present a new algorithm for inferring region information using abstract transformations.

In this section, we assume a finite set $\Typ$ of atomic \emph{types} and use $A,B,C$ to range over atomic types. In addition, we assume a set $\Cls(A) \subseteq \Cls$ of actual classes of an object of type $A$. This allows us to get the set $\Fld(A) \subseteq \Fld$ of fields of (objects of) type $A$. We write $A.f$ to denote the field $f \in \Fld(A)$.

We consider the field typing as a part of an environment; thus, a typing \emph{environment} is a mapping $\Var \cup \Typ \times \Fld \rightharpoonup \cP(\Typ)$ that assigns a variable or a field its possible types. We work with a partial order $\sqsubseteq$ on environments given by $\env \sqsubseteq \env'$ iff $\env(\kappa) \subseteq \env'(\kappa)$ for all $\kappa \in \dom(\env)$. Given an environment $\env$, we write $\vpart{\env} : \Var \rightharpoonup \cP(\Typ)$ and $\fpart{\env} : \Typ \times \Fld \rightharpoonup \cP(\Typ)$ to denote the typings of variables and fields of $\env$ respectively. Given a variable typing $V : \Var \rightharpoonup \cP(\Typ)$ and a field typing $F : \Typ \times \Fld \rightharpoonup \cP(\Typ)$, we write $(V,F)$ to denote the environment combining the typings from $V$ and $F$. In particular, we have $\env = (\vpart{\env},\fpart{\env})$.

We often call a set of atomic types a \emph{type}. We simply write~$A$ to denote the singleton set $\{A\}$ and misuse the disjunction symbol $\vee$ for set unions. The set $\{A,B,C\}$ for example is thus denoted as $A \vee B \vee C$. In particular, we write $\bot$ to denote the empty set of atomic types. For instance, $(x : A,\ A.f : B \vee C)$ is an environment stating that the variable $x$ has type $A$ and the field $f$ of any object of type $A$ can have type $B$ or $C$.

\subsection{Abstract Transformations}
\label{sec:trans:def}

Now we define our notion of abstract transformation which encodes type updates of the variables and fields of a program.

When assigning transformations to statements in the program, the interesting cases are the assignment statements. Consider a statement $x = e$ and its following possible transformations:
\begin{itemize}
\item If $e$ is a constant of type $A$, then the resulting transformation is $[x \tmapsto A]$, meaning that the type of $x$ is $A$.
\item If $e$ is a variable $y$ whose type is unknown yet, then the resulting transformation is $[x \tmapsto y]$, meaning that $x$ has the same type as $y$.
\item If $e$ is a field $y.f$ and the type of $y$ is known to be $A$, then the resulting transformation is $[x \tmapsto A.f]$, meaning that $x$ has the same type as the field $f$ of any object of type $A$.
\item If $e$ is a field $y.f$ and the type of $y$ is unknown, then the resulting transformation is $[x \tmapsto y.f]$, meaning that $x$ has the same type as the field $f$ of any object of the type of $y$.
\end{itemize}
The above cases list four \emph{atomic} kinds of assignment values: atomic type $A$, variable $y$, fields $A.f$ and $y.f$ of a type and a variable. As discussed earlier, we work with access graphs instead of access paths to avoid non-terminating computation. All of above assignment values can be represented using a generalization $b.G$ of access graphs where the base $b$ can also be a type. For instance, the type $A$ is represented by $A.\efgraph$ where $\efgraph$ is the empty field graph, and the field $A.f$ is represented by $A.\fgraph{f}{\emptyset}{f}$. We consider one more possible case of $e$:
\begin{itemize}
\item If $e$ involves some branches and thus has type $B \vee C$, then $x = e$ results in a transformation $[x \tmapsto B \vee C]$, meaning that $x$ has type $B$ or $C$.
\end{itemize}
More generally, the value $v$ of an assignment $x \tmapsto v$ can be the `formal disjunction' of some access graphs $b.G$. These cases bring the following definition of terms to represent assignment values.

\begin{definition}[Atoms and terms]
\label{def:term}
\emph{Atoms} are a generalization of access graphs whose base is either a variable or an atomic type.
We write $b.G$ to denote the atom with base $b \in \Var \cup \Typ$ and field graph $G \in \FG$.

A \emph{term} is simply a set (or a formal disjunction) of atoms. We write $\bot$ to denote the empty term, i.e., the empty set of atoms, and $u \vee v$ to denote the join of terms $u$ and $v$, i.e., the union of the two sets $u,v$ of atoms. Therefore, we have $u \vee \bot = u = \bot \vee u$ for any term $u$.
\end{definition}

When the field graph $G$ is empty, the atom $b.G$ represents a variable or an atomic type. Thus we often omit $G$ and simply write $b$ to denote the atom. If $G = \fgraph{f}{\emptyset}{f}$, that is, a graph consisting of only the singleton field access path $f$, then we may write $b.f$ rather than $b.\fgraph{f}{\emptyset}{f}$.

By definition, each term $u$ has the form $\bigvee^{n}_{i=1} b_i.G_i$ where $u=\bot$ if $n=0$. We \emph{concatenate} a term $u$ with a field graph $G$ by
\[
\textstyle
u.G = (\bigvee^{n}_{i=1} b_i.G_i).G = \bigvee^{n}_{i=1} b_i.(G_i.G)
\]
where the concatenation $G_i.G$ of field graphs has been defined in Section~\ref{sec:access:graph}.

A term is a formal expression that can be instantiated into a concrete type with a given typing environment (see Definition~\ref{def:inst:term}). We denote the set of terms by $\Tm$ and use $u,v,w$ to range over terms.

\begin{definition}[Assignments]
An \emph{assignment} is a pair consists of a variable $x$ and a term $u$, written as $x \tmapsto u$. It means that the type of variable $x$ is the instantiation of the term $u$ w.r.t.\ any typing environment. We call $x$ the \emph{key} of the assignment.
\end{definition}

We want a notion of environment transformation that encodes also the update of field typing. In particular, we choose to work with \emph{weak update} for field typing as in the previous work~\cite{BGH:pointer,EHZ:enforcing,ESX:infinitary,GHL:enforcing}: If a field $f$ of an object of type $A$ is assigned a value of type $B$ and $f$ of another object of the same type $A$ is assigned a value of type $C$, then the field $A.f$ of any object of type $A$ should have a type containing both $B$ and $C$. Therefore, for a statement like $y.f = x$, we cannot give it the assignment $y.f \tmapsto x$ as it expresses that $y.f$ has the same type of~$x$. Instead, we assign it a constraint $y.f \tgeq x$, meaning that the type of the field $f$ of any object of the type of $y$ should be greater than or equal to the type of $x$. If $y$ has type $A$, then the constraint becomes $A.f \tgeq x$. More generally, we define constraints as follows.

\begin{definition}[Constraints]
\label{def:const}
A \emph{constraint} is a pair consisting of a nonempty access graph $b.G$ and a term $u$, written as $b.G \tgeq u$. It means that the type of any field reachable via some path of $b.G$ is greater than or equal to the instantiation of the term $u$ w.r.t.\ any typing environment. We call $b.G$ the key of the constraint.
\end{definition}

Abstract transformations consists of assignments and/or constraints.

\begin{definition}[Abstract transformations]
An \emph{abstract transformation}
\[
[x_1 \tmapsto u_1, \ldots, x_n \tmapsto u_n,\, \kappa_1 \tgeq v_1, \ldots, \kappa_m \tgeq v_m]
\]
is a finite set consisting of assignments $x_i \tmapsto u_i$ and constraints $\kappa_j \tgeq v_j$ such that all the keys are different and $x_i \not= u_i$ for all $i \in \{1,\ldots,n \}$ and $v_j \not= \bot$ for all $j \in \{1,\ldots,m \}$. Let $\sigma$ be the above abstract transformation. We write $\dom(\sigma)$ to denote its \emph{domain}, that is, the set of keys $\{ x_1, \ldots, x_n,\, \kappa_1, \ldots, \kappa_m \}$.
\end{definition}

Let $\FK$ be the set of \emph{keys}, that is, variables and nonempty access graphs. Each abstract transformation $\sigma$ is a representation of a total function from $\FK$ to $\Tm$
\[
\begin{aligned}
\sigma(x) & :=
 \begin{cases}
 u & \text{if } (x \tmapsto u) \in \sigma \\
 x & \text{if } x \not\in \dom(\sigma)
 \end{cases}
& \qquad &
\sigma(\kappa) & :=
 \begin{cases}
 v & \text{if } (\kappa \tgeq v) \in \sigma \\
 \bot & \text{if } \kappa \not\in \dom(\sigma).
 \end{cases}
\end{aligned}
\]
In other words, identity assignments $x \tmapsto x$ and bottom constraints $\kappa \tgeq \bot$ are omitted in abstract transformations. This is because they add no information to the transformations. For instance, if a transformation contains only identity assignments and bottom constraints, then it is instantiated into the identity function on typing environments according to Definition~\ref{def:inst:trans}.

We write $\ATrans$ to denote the set of abstract transformations and use $\sigma,\theta$ to range over abstract transformations in the paper. The empty transformation is denoted as~$[]$, and the one consisting of only bottom assignments $x \tmapsto \bot$ for all variable $x$ is denoted as~$\bot$. As will become clear, $[]$ is the identity environment transformation and $\bot$ the `least' environment transformation.

\begin{example}
Consider again the example in Fig.~\ref{fig:code:trans:env}. The program
\[
\verb|x = y.f; y = new C(); y.f = x|
\]
results in the transformation
\[
[x \tmapsto y.f,\, y \tmapsto C,\, C.f \tgeq y.f].
\]
For \verb|x = y.f|, the type of $\mathtt{y}$ is not known yet and thus it leads to the assignment $x \tmapsto y.f$. In this example, we assume that the type of $\mathtt{new \ C()}$ is some type~$C$ which can be different from class $\mathtt{C}$. Thus $\verb|y = new C()|$ leads to $y \tmapsto C$. The last statement $\verb|y.f = x|$ by itself results in the constraint $y.f \tgeq x$. But because of $x \tmapsto y.f$ and $y \tmapsto C$, the constraint is updated to $C.f \tgeq y.f$ by substituting $y$ in the key $y.f$ by $C$ and the constraint value $x$ by $y.f$. In Section~\ref{sec:trans:comp:join} we will demonstrate how to compose $[x \tmapsto y.f]$, $[y \tmapsto C]$ and $[y.f \tgeq x]$ to get $[x \tmapsto y.f,\, y \tmapsto C,\, C.f \tgeq y.f]$.
\qed
\end{example}

\subsection{Operations on Abstract Transformations}
\label{sec:trans:ops}
Consider again the example from Fig.~\ref{fig:code:trans:env}. In this section, we firstly demonstrate how the transformation $\sigma = [x \tmapsto y.f,\, y \tmapsto C,\, C.f \tgeq y.f]$ updates the environment $\env = (y:A,\, A.f:B)$ to $\sigma(\env) = (x:B,\, y:C,\, A.f:B,\, C.f:B)$. Then we show that abstract transformations can be composed and joined so that we can construct the transformation $\sigma$ for the program from those of its statements.

To begin with, we look into how the type of $x$ is computed in $\sigma(\env)$. There is an assignment $x \tmapsto y.f$ in $\sigma$, meaning that $x$ has the same type as $y.f$. We have to instantiate the term $y.f$ using the typing information given by the input environment $\env$. Because $y$ has type $A$ in $\env$, we instantiate $y.f$ to $A.f$. And because $A.f$ has type $B$ in $\env$, we instantiate $y.f$ further to $B$. Therefore, $x$ has type $B$ in the updated environment $\sigma(\env)$.

We have seen from the above example that we need to instantiate a term to a type according to the environment which we want to update. In particular, we consider how to instantiate an atom $A.\fgraph{h}{E}{t}$. For example, let us instantiate $A.f.g$ according to $(A.f : A \vee B,\, B.g : C,\, C.g : D)$. The goal is to compute the type of the field $g$ of $A.f$. Which field in the environment should be considered, $B.g$ or $C.g$? Because $A.f$ can have type $A$ or $B$, we can \emph{reach} the field $B.g$ but not $C.g$. Therefore, we should instantiate $A.f.g$ only to $C$, i.e., the type of $B.g$.

In the following, we describe how to compute the \emph{reachable fields} from a field $A.h$ via an edge set~$E$ according to the field typing in an environment $\env$. Then, to instantiate $A.\fgraph{h}{E}{t}$ w.r.t.~$\env$, we simply join the types of all fields $B.t$ in $\env$ which are reachable from $A.h$.

\begin{definition}[Reachable fields]
Let $A$ be an atomic type, $h$ a field, $E$ an edge set and $\env$ an environment. We construct the set $\cR(A.h,E,\env) \subseteq \Typ \times \Fld$ of \emph{reachable fields} from $A.h$ via $E$ according to $\env$ as follows:
\begin{itemize}
\item[(1)] Let $\cR(A.h,E,\env) = \{A.h\}$.
\item[(2)] For each $B.f \in \cR(A.h,E,\env)$, let $\cR(A.h,E,\env) = \cR(A.h,E,\env) \cup \cS_{B.f}$, where $\cS_{B.f}$ is the set of \emph{immediate successors} of $B.f$ defined by
\[
\cS_{B.f} := \{ C.g \mid C \in \env(B.f) \text{ and } (f,g) \in E \text{ and } g \in \Fld(C) \}.
\]
\item[(3)] Repeat (2) until $\cR(A.h,E,\env)$ cannot be updated anymore.
\end{itemize}
\end{definition}

Any field $A.f$ is reachable from itself. To compute the other reachable fields from $A.f$, the above algorithm simply gets the immediate successors of $A.f$, and then those of the immediate successors and so on.

\begin{example}
\label{ex:inst:reachable}
Let $\env = (A.f: A \vee B, \, B.g:C)$ and assume $\Fld(A)=\{f,g\}$ and $\Fld(B)=\{g\}$. By definition, we have
\[
\cR(A.f,\emptyset,\env) = \{ A.f \}
\]
because the edge set is empty and thus $f$ has no successors. We have
\[
\cR(A.f,\{(f,g)\},\env) = \{ A.f, A.g, B.g \}
\]
indicating that $A.g$ and $B.g$ are also reachable from $A.f$. That's because $g$ is a successor of $f$ and $g$ is a field of both $A$ and $B$.
\qed
\end{example}

The instantiation $(A.\fgraph{h}{E}{t})[\env] \subseteq \Typ$ is given by the join of $\env(B.t)$ for all reachable fields $B.t \in \cR(A.h,E,\env)$. With this, we can instantiate arbitrary atoms and thus terms.

\begin{definition}[Instantiation of terms]
\label{def:inst:term}
Let $\env$ be an environment. We define the \emph{instantiation} $(b.G)[\env] \subseteq \Typ$ of atom $b.G$ as follows:
\[
\begin{aligned}
A[\env] & := A \\
(A.\fgraph{h}{E}{t})[\env] & := \textstyle \bigvee \left\{ \env(B.t) \mid B.t \in \cR(A.h,E,\env) \right\} \\
(x.G)[\env] & := \textstyle \bigvee \left\{ (A.G)[\env] \mid A \in \env(x) \right\}.
\end{aligned}
\]
The instantiation of a term $u$ is the join of the instantiations of its atoms, i.e.,
\[
\textstyle
u[\env] = (\bigvee^{n}_{i=1} b_i.G_i)[\env] := \bigvee^{n}_{i=1} (b_i.G_i)[\env].
\]
\end{definition}

In the above definition, we assume that if $a \not\in \dom(\env)$ then $\env(a) = \bot$, that is, the empty set of types, where $a$ is a variable $x$ or a field $A.f$. Therefore, we have $x[\env] = \env(x)$ and $(A.f)[\env] = \env(A.f)$.

\begin{example}
\label{ex:inst:term}
Let $\env = (A.f: A \vee B, \, B.g:C)$ and assume $\Fld(A)=\{f,g\}$ and $\Fld(B)=\{g\}$ as in Example~\ref{ex:inst:reachable}. By definition, we have
\[
(A.f)[\env] = \env(A.f) = A \vee B
\]
and
\[
\begin{aligned}
    (A.f.g)[\env]
& = (A.\fgraph{f}{\{(f,g)\}}{g})[\env] \\
& = \textstyle \bigvee \{ \env(X.g) \mid X.g \in \cR(A.f, \{(f,g)\}, \env) \} \\
& = \env(A.g) \vee \env(B.g) \\
& = \bot \vee C = C
\end{aligned}
\]
because from Example~\ref{ex:inst:reachable} we know both $A.g$ and $B.g$ are reachable from $A.f$.
\qed
\end{example}

Lastly, we instantiate abstract transformations to endofunctions on typing environments. The type of a variable is computed by instantiated the its assigned term in the transformation, as discussed above. To compute the types of fields, we solve the constraints using a fixed-point algorithm.

\begin{definition}[Instantiation of abstract transformations]
\label{def:inst:trans}
Let $\sigma$ be an abstract transformation. We define an endofunction $\varphi_\sigma$ on environments by
\[
\begin{aligned}
\varphi_\sigma(\env)(x) & := \begin{cases} u[\env] & \text{if } (x \tmapsto u) \in \sigma \\ \env(x) & \text{otherwise} \end{cases} \\
\varphi_\sigma(\env)(A.f) & := \env(A.f) \vee \textstyle \bigvee \{ u[\env] \mid (b.\fgraph{h}{E}{f} \tgeq u) \in \sigma \text{ and } \\
& \hspace{116pt} B \in b[\env] \text{ and } A.f \in \cR(B.h,E,\env) \}.
\end{aligned}
\]
Let $\env$ be an environment. We define an environment $\sigma(\env)$ by
\begin{enumerate}
\item Let $\env' = \env$.
\item \label{comp:item:update} Let $\env'' = \varphi_\sigma(\vpart{\env}, \fpart{\env'})$, where $(\vpart{\env}, \fpart{\env'})$ is the environment obtained by combining the variable typing of $\env$ and the field typing of $\env'$.
\item If $\env' \not= \env''$, then let $\env' = \env''$ and go back to Step~\ref{comp:item:update}. Otherwise, let $\sigma(\env) = \env'$.
\end{enumerate}
This procedure results in an environment transformation mapping $\env$ to $\sigma(\env)$.
\end{definition}

To instantiate an abstract transformation $\sigma$, we use the above fixed-point algorithm to solve the constraints for field typings in $\sigma$. What crucial is the function $\varphi_\sigma$ that updates the environment in each iteration towards the fixed point. If $(x \tmapsto u) \in \sigma$, then $\varphi_\sigma(\env)$ assigns $x$ to the instantiation $u[\env]$; otherwise, $x$ is assigned the type as claimed in $\env$. Because of weak update for field typing, $\varphi_\sigma(\env)$ assigns a field $A.f$ to the join containing its previous type $\env(A.f)$ given by the environment and the instantiations of constraint values from whose keys the field $A.f$ can reach.

Note that, in each iteration towards the fixed point, the input of $\varphi_\sigma$ consists of the variable typing $\vpart{\env}$ from the original environment $\env$ and the field typing $\fpart{\env'}$ from the result $\env'$ of the previous iteration. This is because types of variables and of fields are updated in different manners. We `accumulate' the field typing by feeding $\varphi_\sigma$ with the field typing from the previous iteration due to weak update as explained earlier. However, variable typing is not updated in this way. For instance, consider the code \verb|x = y; y = new C()|, resulting in the transformation $\sigma = [x \tmapsto y,\, y \tmapsto C]$, and the environment $\env = (y:A)$. We have $\varphi_\sigma(\env) = (x:A,\, y:C)$ which gives the correct type to $x$, because~$x$ should have the same type of $y$ before the assignment \verb|y = new C()| which is~$A$. But applying $\varphi_\sigma$ to the updated environment would give $x$ type $C$.

By definition, the empty transformation $[]$ is identity on environments. For a more interesting example, we consider the transformation from Fig.~\ref{fig:code:trans:env}.

\begin{example}
\label{ex:inst:trans}
Recall the abstract transformation from Fig.~\ref{fig:code:trans:env}:
\[
\sigma = [x \tmapsto y.f,\, y \tmapsto C,\, C.f \tgeq y.f].
\]
Let $\env = (y : A,\, A.f : B)$. The variables and fields that should appear in the updated environment $\sigma(\env)$ consist of $x$, $y$, $A.f$ and $C.f$. By definition, we have
\[
\begin{aligned}
\varphi_\sigma(\env)(x) & = (y.f)[\env] = B \\
\varphi_\sigma(\env)(y) & = C[\env] = C \\
\varphi_\sigma(\env)(A.f) & = \env(A.f) \vee \bot = B \\
\varphi_\sigma(\env)(C.f) & = \env(C.f) \vee (y.f)[\env] = \bot \vee B = B. \\
\end{aligned}
\]
Because $\varphi_\sigma(\env) = \varphi_\sigma(\vpart{\env},\fpart{\varphi_\sigma(\env)})$, we reach the fixed point and get the updated environment $\sigma(\env) = \varphi_\sigma(\env) = (x:B,\, y:C,\, A.f:B,\, C.f:B)$.
\qed
\end{example}

As explained in the next section, each statement of a program can be assigned a transformation indicating the assignment or constraint of the involved type. To combine them into one transformation that summarizes the type updates of the whole program, the operations of composition and join for transformations are needed. Due to the lack of space, we characterize these operations in the following theorems. Details of their (non-surprising) constructions are available in Appendix~\ref{sec:trans:comp:join}.

\begin{theorem}[Composition of abstract transformations]
For any abstract transformations $\sigma$ and $\theta$, we can construct an abstract transformation $\delta$ such that $\sigma(\theta(\env)) \sqsubseteq \delta(\env)$. We write $\sigma\theta$ to denote $\delta$ and call it the \emph{composition} of $\sigma$ and $\theta$. Moreover, we have $\sigma[]=\sigma=[]\sigma$ for any transformation $\sigma$, where $[]$ is the empty transformation.
\qed
\end{theorem}

The difficult part of the work is to come up with the right notion of abstract transformation that supports composition. But the construction of composition and its correctness are then straightforward. The idea to compose our abstract transformations is similar to the one for substitutions (see e.g.~\cite[\S2.1]{DBLP:books/el/RV01/BaaderS01}). And its correctness can be proved with a standard inductive argument on the length of the abstract transformation.

Note that we have only $\sigma(\theta(\env)) \sqsubseteq (\sigma\theta)(\env)$ where $\sqsubseteq$ is the ordering on environments defined pointwisely. This is because $\sigma\theta$ involves concatenation of field graphs which causes the over approximation. For instance, concatenating a singleton path $f$ with itself does not give the path $f.f$. Instead it results in the field graph $\fgraph{f}{\{(f,f)\}}{f}$ which represents all the paths consisting of $f$ with length greater than 1. However, the type inference algorithm presented in next section is still sound. It may give a less precise type to the program.

\begin{theorem}[Join of abstract transformations]
For any abstract transformations $\sigma$ and $\theta$, we can construct an abstract transformation $\delta$ such that $\sigma(\env) \sqcup \theta(\env) \sqsubseteq \delta(\env)$. We write $\sigma \vee \theta$ to denote $\delta$ and call it the \emph{join} of $\sigma$ and $\theta$. Moreover, we have $\sigma \vee \theta = \theta \vee \sigma$ and $\sigma \vee \bot =\sigma$ for all $\sigma$ and $\theta$, where $\bot$ is the bottom transformation that assigns all variables to the bottom type.
\qed
\end{theorem}

The join $\sigma \vee \theta$ is constructed componentwise using the join operator on terms. It does not preserve fixed points and we have only $\sigma(\env) \sqcup \theta(\env) \sqsubseteq (\sigma \vee \theta)(\env)$. As discussed above, this causes no harm to the soundness of the type inference.

\section{Type Inference via Abstract Transformations}
\label{sec:type:infer}

In this section, we demonstrate how to infer the type of an FJ program using abstract transformations. The idea is to use abstract transformations to capture the flow information of a program, which leads to a more efficient type inference algorithm. As an example, we work with the region type system of Beringer et al.~\cite{BGH:pointer}. Our inference algorithm firstly computes an abstract transformation for each method of the program, and then uses them for the type inference rather than analyzing the method bodies.

\subsection{Region Type System}
\label{sec:region}

We briefly recall the region type system of Beringer et al.~\cite{BGH:pointer}.

A \emph{region} represents a property of a value such as its provenance information. In this paper, we use the following definition of regions: 
\[ 
  \begin{aligned}
    \Reg \ni r, s\ ::=\ \Null \Mid \Createdat{\ell}
  \end{aligned} 
\]
The region $\Null$ contains only the value $\itnull$. The region $\Createdat{\ell}$ contains all references to objects that were created by an expression of the form $\new{\ell}{C}$. This region allows us to track where in the program an object originates. One can use a richer definition of regions to capture other properties of interest such as taintedness~\cite{EHZ:enforcing}. We keep it simple here because we focus on the type inference.

Region type information is complementary to FJ type information and can be captured without repeating the FJ type system. Therefore, we directly work with region types rather than refining FJ types as in the original system~\cite[Section~3]{BGH:pointer}.

As for FJ, we need a class table to record the region types of methods and fields. This is needed to formulate typing rules for method call and field access. A \emph{class table} $(F,M)$ consists of
\begin{itemize}
\item a \emph{field typing} $F : \Cls \times \Reg \times \Fld \rightharpoonup \cP(\Reg)$ that assigns to each class $C$, region $r$ and field $f \in \fields(C)$ a set $F(C,r,f)$ of regions of the field~$f$, and
\item a \emph{method typing} $M : \Cls \times \Reg \times \Mtd \times \Reg^* \rightharpoonup \cP(\Reg)$ that assigns to each class $C$, region $r$, method $m \in \methods\left(C\right)$ and sequence $\bar{s}$ of regions of $m$'s formal arguments a set $M(C,r,m,\bar{s})$ of regions of the method $m$.
\end{itemize}
The typing functions are required to be well-formed, which reflects the subtyping properties of FJ. See \cite[Definition 4.2]{ESX:infinitary} for the details.

Typing judgments take the form $\Gamma \vdash e : R$, where $\Gamma: \Var \rightharpoonup \Reg$ is a typing environment for variables, $e \in \Expr$ a term expression and $R \subseteq \Reg$ a set of regions. The typing rules are listed in Figure~\ref{fig:typing}. For instance, the \textsc{call} rule looks up the method typing $M$ for all possible regions where the object~$x$ and arguments~$\bar{y}$ may reside and joins the matched entries as the return type of the method invocation $x.m(\bar{y})$.

\begin{figure}[!h]
\[
\inferrule*[left={var}]
{ }
{\Gamma,\, x\of R  \vdash x : R}
\qquad
\inferrule*[left={null}]
{ }
{\Gamma \vdash \ttnull : \Null}
\qquad
\inferrule*[left={new}]
{ }
{\Gamma \vdash \new{\ell}{C} : \Createdat{\ell}}
\]
\[
\inferrule*[left={sub}]
{\Gamma \vdash e : R\\
 R \subseteq R'}
{\Gamma \vdash e : R'}
\qquad
\inferrule*[left={cast}]
{\Gamma \vdash e : R}
{\Gamma \vdash \cast{D}{e} : R}
\]
\[
\inferrule*[left={if}]
{\Gamma,\, x\of R \cap S,\, y\of R \cap S \vdash e_1 : T_1 \\
 \Gamma,\, x\of R,\, y\of S \vdash e_2 : T_2}
{\Gamma,\, x\of R,\, y\of S \vdash \ttifthenelse{x}{y}{e_1}{e_2} : T_1 \cup T_2}
\]
\[
\inferrule*[left={let}]
{\Gamma \vdash e_1 : R\\
 \Gamma,\, x\of R \vdash e_2 : T}
{\Gamma \vdash \letin{x}{e_1}{e_2} : T}
\qquad
\inferrule*[left={call}]
{\textstyle T = \bigcup \{ M(C, r , m, \bar{s}) \mid r \in R, \bar{s} \in \bar{S} \}}
{\Gamma,\, x\of R ,\, \bar{y}\of \bar{S} \vdash x^C.m(\bar{y}) : T}
\]
\[
\inferrule*[left={get}]
{\textstyle T = \bigcup \{ F(C,r,f) \mid r \in R \}}
{\Gamma,\, x\of R \vdash x^C.f : T}
\qquad
\inferrule*[left={set}]
{\forall r \, \mathord{\in}\, R.\ S \subseteq F(C,r,f)}
{\Gamma,\, x\of R,\, y\of S \vdash x^C.f := y : S}
\]
\caption{The region type system of Beringer et al.~\cite{BGH:pointer}}
\label{fig:typing}
\end{figure}

An FJ program $(\prec, \fields, \methods,$ $\mtable)$ is \emph{well-typed} w.r.t.\ a class table $(F,M)$ if for any $(C,r,m,\bar{s})$ with $M(C,r,m,\bar{s}) = R$ and $\mtable\left(C,m\right)=(\bar{x},e)$, the typing judgment $\mathtt{this} \of r,\, \bar{x} \of \bar{s} \vdash e : R$ is derivable. A soundness theorem has been proved in~\cite[Theorem~1]{BGH:pointer}, stating that, for any expression $e$ in a well-typed FJ program with respect to a class table $(F,M)$, if $e$ evaluates to some value $v$ and $e$ has type $R$, then $v$ is in some region in $R$.

\subsection{Inferring Region Types via Abstract Transformations}
\label{sec:infer:regions}

Let an FJ program $P$ be given. Now we introduce an algorithm to construct a class table $(F,M)$ with respect to which $P$ is well-typed. As mentioned above, our approach is based on abstract transformations. From now on, the atomic types we are working with are the regions, i.e., take $\Typ = \Reg$ for the development of abstract transformations.

We firstly compute an abstract transformation $\sigma$ and a term $t$ for each FJ expression $e$. The transformation $\sigma$ encodes the type updates of the variables and fields in $e$, while the term $t$ pre-calculates type of $e$. Once we are given a typing environment $\env$, we update it using $\sigma$ and then instantiate $t$ with the updated environment to compute the type of $e$, i.e., $e$ has type $t[\sigma(\env)]$.

For this, we define the following operations on pairs of abstract transformations and terms: Let $(\sigma, s), (\theta, t) \in \ATrans \times \Tm$.
\begin{itemize}
\item \emph{Composition}: We define $(\sigma,s)\theta := (\sigma\theta,s\theta)$, where $\sigma\theta$ is the composition of transformations, and $s\theta$ is term substitution.
\item \emph{Join}: We define $(\sigma,s) \vee (\theta,t) := (\sigma \vee \theta, s \vee t)$.
\end{itemize}

Suppose we have a function $T : \Cls \times \Mtd \to \ATrans \times \Tm$, called an \emph{abstract method table}, that assigns an abstract transformation and a term to each method. The transformations capture the type updates for the method and the term will be instantiated to the type for the method. Then we define a pair $\sub{e} : \ATrans \times \Tm$ by induction on the FJ expression $e$:
\[
\begin{aligned}
\sub{x} & := ([], x) \\
\sub{\ttifthenelse{x}{y}{e_1}{e_2}} & := \sub{e_1} \vee \sub{e_2} \\
\sub{\letin{x}{e_1}{e_2}} & := \sub{e_2} ([x \tmapsto t]\theta) \quad \text{where } (\theta,t) = \sub{e_1} \\
\sub{\ttnull} & := ([], \Null) \\
\sub{\new{\ell}{C}} & := ([], \Createdat{\ell}) \\
\sub{\cast{D}{e}} & := \sub{e} \\
\sub{x.f} & := ([], x.f) \\
\sub{x.f := y} & := ([x.f \tgeq y], y) \\
\sub{x^C.m(\bar{y})} & := T(C,m) [\this \tmapsto x, \args(C,m) \tmapsto \bar{y}]
\end{aligned}
\]
where $\args(C,m)$ denotes the arguments of $m$, i.e., if $\mtable(C,m) = (\bar{x},e)$ then $\args(C,m)=\bar{x}$. 

Given an FJ program, we compute an abstract method table $T$ as follows:
\begin{enumerate}
\item Initialize $T$ with $T(C,m) = ([],\bot)$ for all $m \in \methods(C)$, where $\bot$ is the empty term, i.e., the empty set of atoms.
\item\label{step:T} For each method, compute an abstract transformation and a term for its body, and then update the corresponding entry in~$T$. Specifically, for each $m \in \methods(C)$ with $(\bar{x},e) = \mtable(C,m)$, let $T(C,m) = T(C,m) \vee \sub{e}$.
\item\label{step:T:closure} Close $T$ under the subclass relation, i.e., let $T(C,m) = T(C,m) \vee T(D,m)$ if $D$ is a subclass of $C$.
\item Repeat steps \ref{step:T} and \ref{step:T:closure} until no more update of $T$ is possible.
\end{enumerate}

After computing the table~$T$, we compute a class table $(F,M)$ as follows:
\begin{itemize}
\item[(a)] Initialize $F$ and $M$ with the least type, i.e.,\ the empty set of regions.
\item[(b)] Use $T$ to update the entries in $F$ and $M$. Specifically, for each $C,r,m,\bar{s}$ with $(\sigma,u) = T(C,m)$, we update the environment and get $\env = \sigma(\Gamma,F)$ where $\Gamma = \this \of r,\, \args(C,m) \of \bar{s}$. Then we update the class table by taking $F = F \vee \fpart{\env}$ and $M(C,r,m,\bar{s}) = M(C,r,m,\bar{s}) \vee u[\env]$.
\item[(c)] Ensure that $(F,M)$ are well-formed. For instance, if $D$ is a subclass of $C$, then both $F(C,r,f)$ and $F(D,r,f)$ are set to their join.
\item[(d)] Repeat steps (b) and (c) until no more update of $F$ and $M$ is possible.
\end{itemize}

To summarize, the inference algorithm has two steps. It firstly computes an abstract method table $T$. The abstract transformations in $T$ capture the flow information of each method. This step is similar to constraint generation and preprocessing such as simplification or closure in the constraint-based type inference~\cite{10.1007/978-3-642-25318-8_10,10.5555/646158.679874}. Then it computes the class table $(F,M)$ by instantiating the abstract transformations in $T$. This step solves the constraints collected in $T$ via a least fixed-point argument.

The inference algorithm in the previous work~\cite[Appendix F]{EHZ:enforcing} analyzes the same method body multiple times when the method is fed with arguments of different types at different invocations. Our algorithm instead uses the abstract transformation stored in~$T$ to infer the types of different method calls. Therefore, it can effectively enhance the efficiency of analysis especially when the analyzed program contains many method invocations with arguments of different types. Appendix~\ref{sec:infer:example} has an example demonstrating how to compute and use $T$ to analyze invocations of the same method.

Lastly, the computed class table $(F,M)$ reveals the region information of the program in the following sense.

\begin{theorem}[Correctness of type inference]
Let $P$ be an FJ program. The above algorithm gives a class table $(F,M)$ with respect to which $P$ is well-typed. In particular, for any $C,r,m,\bar{s}$ with $M(C,r,m,\bar{s}) = R$ and for any $x:r$ and $\bar{y} : \bar{s}$, if $x.m(\bar{y})$ evaluates to a value $v$, then $v$ resides in some region in $R$.
\qed
\end{theorem}

The second part of the above theorem is a corollary of the soundness result~\cite[Theorem~1]{BGH:pointer}. It states that the type of each method computed by $M$ is correct. We sketch the proof of the first claim that the program is well-typed with respect to the class table $(F,M)$ given by our algorithm. Because the typings $F$ and $M$ are computed by the abstract transformations of the table $T$, we only need to prove that these abstract transformations compute types greater than the ones from typing derivation. More precisely, we need to prove
\[
\text{if } \Gamma \vdash e : R \text{ then } R \subseteq \mathsf{Type}(\sub{e}(\Gamma))
\]
where $\mathsf{Type}(\sub{e}(\Gamma))$ is the type of the expression~$e$ obtained by firstly applying the transformation component of $\sub{e}$ to $\Gamma$ to obtained an updated environment and then instantiating the term component with the updated environment. The above statement can be proved by induction on the length of typing derivation as usual, because the definition of $\sub{e}$ reflects the typing rules.

\section{Conclusion, Implementation and Discussion}
\label{sec:conclusion}

In this paper, we develop a theory of abstract transformations to capture type changes in programs. The elements of an abstract transformation can be viewed as equality and subtyping constraints. In particular, we work with access graphs when defining these constraints. Access graphs provide a finite representation of field access paths and thus ensure the termination of the procedure to compute abstract transformations for the program. We instantiate abstract transformations to endofunctions on typing environments to compute the types of the program, which solves the constraints in the abstract transformations. As an example, we work with the region type system of Beringer et al.~\cite{BGH:pointer} to demonstrate how to use our inference algorithm based on abstract transformations to compute region information of Featherweight Java programs. The advantage is that the code of a method is analyzed only once even when it is invoked with arguments of different region types in multiple occasions of the program.

We have a prototype implementation of the type inference algorithm using the Soot framework~\cite{soot}. It takes a Java bytecode program as input and computes the region types of the program. The implementation of abstract transformations and their operations follows the definitions in this paper. The function $\sub{-}$ computing an abstract transformation and a type term for FJ expressions becomes a forward flow analysis for the control flow graphs of the program. In particular, it has a flow-through method that computes an abstract transformation for each node in the control flow graph and then concatenates it with the one generated from the previous nodes. Then a fixed point procedure is implemented to compute an abstract transformation for each method in the program using the flow analysis. Lastly, the generated abstract transformations are instantiated to compute the types of the methods in the program. The prototype implementation is available at our GitHub repository\footnote{\url{https://github.com/cj-xu/AbstractTransformation}}.

Region types can make the analysis of trace properties more precise~\cite{EHZ:enforcing,ESX:infinitary}. By extending region type systems with effect annotations to give information about possible event traces of the program, a method invocation $x.m(\bar{y})$ can have different effects for $x,\bar{y}$ in different regions. Our approach can be extended to reason also such region-sensitive trace effects. Our idea is to make the abstract method table $T$ to compute also a formal expression capturing the information of traces and method calls. For example, consider the following FJ program
\[
\tto(a);\ x.f(\bar{y});\ \tto(b);\ x.g(\bar{z})
\]
where $\tto(a)$ is a primitive method that emits the event $a$. We can assign it a formal expression
\[
\{a\} \cdot X_{(x,f,\bar{y})} \cdot \{b\} \cdot X_{(x,g,\bar{z})}
\]
meaning that any trace generated by the program starts with the event $a$, followed by a trace generated by the method call $x.f(\bar{y})$ and then the event $b$ and a trace generated by $x.g(\bar{z})$. Here $x,\bar{y},\bar{z}$ are variables and can be instantiated to region types with a given environment; thus, the formal expression can be instantiated such that it contains only variables like $X_{(r,f,\bar{s})}$ for the effect of the method call of $f$ for an object in region $r$ with arguments in regions $\bar{s}$. For each method in a calling context, we use its abstract transformation to update the environment and then use the updated environment to instantiate its call expression. In this way, we obtained a set of call expressions, one for each method invocation in a calling context. Then we can use a least fixed point algorithm to compute the trace effect of each method from these call expressions. Currently we are still tackling the details to develop such a compositional algorithm for inferring region-sensitive trace effects.

\subsubsection{Acknowledgements} We thank Fredrick Nordvall Forsberg for the fruitful discussion on this work and the anonymous reviewers for their valuable comments and suggestions on the paper and its accompanying artifact.

\bibliographystyle{splncs04}
\bibliography{GFbib}

\appendix

\section{Composition and Join of Abstract Transformations}
\label{sec:trans:comp:join}

We explain how the operations of composition and join for abstract transformations are defined. They are essential for modeling type updates of programs.

When composing an abstract transformation $\sigma$ with another one $\theta$, we need to substitute each term in $\sigma$ according to $\theta$.

\begin{definition}[Substitution of terms]
\label{def:subst:term}
We \emph{substitute} an atom $b.G$ according to a given abstract transformation $\theta$ to obtain a term $(b.G)\theta \in \Tm$ as follows:
\[
\begin{aligned}
(b.G)\theta & :=
  \begin{cases}
  b.G & \text{$b \not\in \dom(\theta)$}\\
  u.G & \text{if $(b \tmapsto u) \in \theta$}.
  \end{cases}
\end{aligned}
\]
The \emph{substitution} of a term is the join of the substitutions of its atoms, i.e.,
\[
u\theta = \textstyle (\bigvee^{n}_{i=1} b_i.G_i)\theta := \textstyle \bigvee^{n}_{i=1} (b_i.G_i)\theta.
\]
\end{definition}

In other words, if $\theta$ contains an assignment $x \tmapsto v$, then every occurrence of~$x$ in~$u$ is `replaced' by~$v$ using field graph composition to obtain the term $u\theta$. For instance, we have $x\theta = v$ if $(x \tmapsto v) \in \theta$, while $A\theta = A$ because $A$ as a type cannot be in $\dom(\theta)$. In particular, substituting with the empty transformation $[]$ makes no change to the term, i.e., $u[]=u$ for all terms $u$.

For assignment $(x \tmapsto u) \in \sigma$, we only need to substitute its value $u$ by $\theta$ when constructing the composition $\sigma\theta$. However, for constraint $(y.G \tgeq v) \in \sigma$, we need to substitute also the variable $y$ in its key according to $\theta$. Because $\theta$ may assign $y$ to an arbitrary term $w$, the substitution $w.G$ of the key $y.G$ may not be a key but instead a set of keys. Therefore, the substitution $(y.G \tgeq v)\theta$ is a set of constraints mapping each new key to $v\theta$.

\begin{definition}[Substitution of constraints]
\label{def:subst:const}
We \emph{substitute} a constraint ${\kappa \tgeq u}$ according to an abstract transformation $\theta$ and get a set $(\kappa \tgeq u)\theta$ of constraints by
\[
(\kappa \tgeq u)\theta := \{ \alpha_j \tgeq u\theta  \mid \kappa\theta = \textstyle \bigvee^{n}_{i=1} \alpha_i \text{ and } 1 \leq j \leq n \}.
\]
\end{definition}

Note that substitution is required only for constraints whose key involves a variable. For instance, we have $(A.G \tgeq u)\theta = \{ A.G \tgeq u \}$ because $(A.G)\theta = A.G$. Moreover, substitution of constraints can result in the empty set. For example, we have $(x.G \tgeq u)\theta = \emptyset$ if $(x \tmapsto \bot) \in \theta$. Lastly, substituting with the empty transformation $[]$ makes no change to the constraint, i.e., $(\kappa \tgeq u)[]=\{ \kappa \tgeq u \}$.

Now we define the composition $\sigma\theta$ where the type updates of the transformation $\theta$ are considered to be performed first.

\begin{definition}[Composition of abstract transformations]
\label{def:comp}
Given two abstract transformations
\[
\begin{aligned}
\sigma & = [x_1 \tmapsto u_1, \ldots, x_n \tmapsto u_n,\, \alpha_1 \tgeq v_1, \ldots, \alpha_m \tgeq v_m] \\
\theta & = [y_1 \tmapsto s_1, \ldots, y_k \tmapsto s_k,\, \beta_1 \tgeq t_1, \ldots, \beta_l \tgeq t_l]
\end{aligned}
\]
we compose them to get an abstraction transformation $\sigma\theta$ as follows:
\begin{enumerate}
\item \label{def:item:comp:subst} Build the following set of assignments and constraints
 \[
 \Delta = \{ x_1 \tmapsto u_1\theta, \ldots, x_n \tmapsto u_n\theta \}
          \cup (\alpha_1 \tgeq v_1)\theta \cup \cdots \cup (\alpha_m \tgeq v_m)\theta
          \cup \theta.
 \]
\item \label{def:item:comp:del} Remove from $\Delta$
  \begin{itemize}
  \item[i.] all the identity assignments $x \tmapsto x$,
  \item[ii.] all the bottom constraints $\kappa \tgeq \bot$, and
  \item[iii.] any assignment $(y \tmapsto s) \in \theta$ such that $y = x_i$ for some $i \in \{ 1, \ldots, n \}$.
  \end{itemize}
\item \label{def:item:comp:merg} Remove the constraints $(\kappa \tgeq w_1), \ldots, (\kappa \tgeq w_o) \in \Delta$ with the same key and add the merged one $(\kappa \tgeq w_1 \vee \cdots \vee w_o)$ into $\Delta$.
\end{enumerate}
The resulting set of assignments and constraints is an abstract transformation.
\end{definition}

The idea of the algorithm to compose abstract transformations $\sigma$ and $\theta$ is the following: We firstly substitute all the assignments and constraints in $\sigma$ according to $\theta$. The substitution result together with the elements of $\theta$ forms a ``pseudo transformation'' $\Delta$ (Step~\ref{def:item:comp:subst}) which needs to be cleaned up. Specifically, all the identity assignments and bottom constraints should be removed (Steps \ref{def:item:comp:del}.i and \ref{def:item:comp:del}.ii), and the duplicate keys should be taken care of. If there are multiple assignments with the same variable key, then one of them must come from $\theta$ and it should be removed (Step~\ref{def:item:comp:del}.iii). This is because the assignments in~$\sigma$ are performed after and thus overwrite those in $\theta$. For instance, the code \verb|x = y; x = z| leads to the composition $[x \tmapsto z] [x \tmapsto y]$ (in ``backward'' order); the resulting transformation is $[x \tmapsto z]$ as the earlier assignment $x \tmapsto y$ is overwritten. If there are multiple constrains with the same field key, then their values should be joined (Step~\ref{def:item:comp:merg}). This is due to the \emph{weak update} for field typing. For instance, if both $x$ and $y$ refer to some objects of type $A$, then the code \verb|x.f = z; y.f = z'| leads to the composition $[A.f \tgeq z'][A.f \tgeq z]$ which evaluates to $[A.f \tgeq z \vee z']$.

\begin{example}
We consider again the example from Fig.~\ref{fig:code:trans:env}. The statements of the program and their assigned abstract transformations are given below:
\[
\begin{array}{lll}
\verb|x = y.f;| & \qquad &
 [x \tmapsto y.f]\\
\verb|y = new C();| &&
 [y \tmapsto C]\\
\verb|y.f = x| &&
 [y.f \tgeq x]\\
\end{array}
\]
We compose the first two transformations and have
\[
  \theta
= [y \tmapsto C][x \tmapsto y.f]
= [y \tmapsto C,\, x \tmapsto y.f].
\]
This easy composition does not involve any substitution. To compose the last transformation, we need to substitute its constraint according to $\theta$:
\[
  (y.f \tgeq x)\theta
= \{ \kappa \tgeq x\theta \mid \kappa \in (y.f){\theta} \}
= \{ C.f \tgeq y.f \}
\]
because $(y.f){\theta} = C.f$ and $x\theta = y.f$. Finally, the composition of the three transformations is
\[
  [y.f \tgeq x]([y \tmapsto C][x \tmapsto y.f])
= [x \tmapsto y.f,\, y \tmapsto C,\, C.f \tgeq y.f]
\]
which does not require cleaning up.
\qed
\end{example}

It is clear that (pre- or post-)composing with the empty transformation $[]$ is identity, i.e., $[]\sigma = \sigma = \sigma[]$ for any transformation $\sigma$.

Given two abstract transformations, we join them componentwise as follows.

\begin{definition}[Join of abstract transformations]
\label{def:join}
The join $\sigma \vee \theta$ consists of non-identity assignments $x \tmapsto \sigma(x) \vee \theta(x)$ and constrains $\kappa \tgeq \sigma(\kappa) \vee \theta(\kappa)$ for $x,\kappa \in \dom(\sigma) \cup \dom(\theta)$.
\end{definition}

The above definition uses the convention $\sigma(x)=x$ and $\sigma(\kappa) = \bot$ for variables $x \not\in \dom(\sigma)$ and nonempty access graphs $\kappa \not\in \dom(\sigma)$ introduced in Section~\ref{sec:trans:def}.

Because the join operation of terms is commutative, so is the one of abstract transformations. Moreover, we have $\sigma \vee \bot = \sigma = \bot \vee \sigma$ for all transformations $\sigma$, where $\bot$ is the bottom transformation which consists of assignments $x \tmapsto \bot$ for all variables $x$.

\section{An Example of Inferring Region Types}
\label{sec:infer:example}
We present an example to illustrate how to infer the region type of a program using the algorithm introduced in Section~\ref{sec:infer:regions}.

Consider the following Java code taken from~\cite{ESX:infinitary}.
\vspace{-12pt}
\begin{center}
\begin{minipage}[t]{.4\textwidth}
\begin{lstlisting}
class Node {
 Node next;
 Node last() {
  if (next == null) {
   return this;
  } else {
   return next.last();
  }
 }
}
\end{lstlisting}
\end{minipage}
\hspace{10pt}
\begin{minipage}[t]{.4\textwidth}
\begin{lstlisting}
class Test {
 Node linear() {
  Node x = new$^{\ell_1}$ Node();
  Node y = new$^{\ell_2}$ Node();
  y.next = x;
  return y.last();
 }
 Node cyclic() {
  Node z = new$^{\ell_3}$ Node();
  z.next = z;
  return z.last();
 }
}
\end{lstlisting}
\end{minipage}
\end{center}
\vspace{-4pt}
The two methods in $\mathtt{Test}$ create a linear linked list and a cyclic one. We illustrate how to compute the entry $T(\Node,\ttlast)$ and then use it to infer the types of the methods in $\mathtt{Test}$.

We initialize $T(\Node,\ttlast)=([],\bot)$ and then increase it using $\sub{e_{\ttlast}}$ where $e_{\ttlast}$ is the method body of $\ttlast$:
\begin{itemize}
\item In the first iteration, we have
 \[
 \begin{aligned}
 \sub{e_{\ttlast}} & = ([],\this) \vee (T(\Node,\ttlast)[\this \tmapsto \this.\ttnext]) \\
 & = ([],\this) \vee ([\this \tmapsto \this.\ttnext], \bot) \\
 & = ([\this \tmapsto \this \vee \this.\ttnext], \this) \\[2pt]
 T(\Node,\ttlast) & = T(\Node,\ttlast) \vee \sub{e_{\ttlast}} \\
 & = ([], \bot) \vee ([\this \tmapsto \this \vee \this.\ttnext], \this) \\
 & = ([\this \tmapsto \this \vee \this.\ttnext], \this).
 \end{aligned}
 \]
\item In the second iteration, we have
 \[
 \begin{aligned}
 \sub{e_{\ttlast}} = \cdots & = ([\this \tmapsto \this \vee \this.\ttnext \vee \this.\overline{\ttnext}],\\
 & \hspace{18pt} \this.\ttnext) \\[2pt]
 T(\Node,\ttlast) = \cdots & = ([\this \tmapsto \this \vee \this.\ttnext \vee \this.\overline{\ttnext}],\\
 & \hspace{18pt} \this \vee \this.\ttnext)
 \end{aligned}
 \]
 where $\this.\overline{\ttnext} = \this.\fgraph{\ttnext}{\{(\ttnext,\ttnext)\}}{\ttnext}$ is the access graph consisting of all paths with at least two $\ttnext$'s.
\item In the third iteration, we have
 \[
 \begin{aligned}
 \sub{e_{\ttlast}} = \cdots & = ([\this \tmapsto \this \vee \this.\ttnext \vee \this.\overline{\ttnext}],\\
 & \hspace{18pt} \this.\ttnext) \\[2pt]
 T(\Node,\ttlast) = \cdots & = ([\this \tmapsto \this \vee \this.\ttnext \vee \this.\overline{\ttnext}],\\
 & \hspace{18pt} \this \vee \this.\ttnext \vee \this.\overline{\ttnext}).
 \end{aligned}
 \]
\end{itemize}
Then we reach the fixed point and have computed $T(\Node,\ttlast)$.

Let $r_i = \Createdat{\ell_i}$ be the regions for objects created at $\ell_1, \ell_2, \ell_3$. We compute the return types of $\linear$ and $\cyclic$, that is, the possible regions where the returned object resides.
The $\linear$ method returns \verb|y.last()| resulting in $T(\Node,\ttlast)[\this \tmapsto \mathtt{y}]$ whose second component is a term
\[
t = \mathtt{y} \vee \mathtt{y}.\ttnext \vee \mathtt{y}.\overline{\ttnext}.
\]
It is obvious that then environment at the return site is
\[
\env = (\mathtt{x}:r_1,\, \mathtt{y}:r_2,\, r_1.\ttnext : \Null,\, r_2.\ttnext : r_1).
\]
The return type of $\linear$ is the instantiation
\[
t[\env] = r_1 \vee r_2.
\]
With a similar computation, we have that the return type of $\cyclic$ is $r_3$.

\end{document}